\documentclass[a4paper,twocolumn,11pt,accepted=2025-01-03]{quantumarticle}
\pdfoutput=1
\usepackage{amsmath,epsfig,mathrsfs,graphicx,amssymb,color}
\usepackage[T1]{fontenc}
\usepackage{t1enc}
\usepackage{hyperref}
\usepackage{array}
\usepackage{booktabs}
\usepackage{yquant} 
\usetikzlibrary{fit, quotes}
\useyquantlanguage{groups}
\usepackage{physics}

\def\be#1\ee{\begin{equation}#1\end{equation}}
\newcommand{\ba}{\begin{eqnarray} }
\newcommand{\ea}{\end{eqnarray} }

\def\mb{\begin{pmatrix}}
\def\me{\end{pmatrix}}
\def\be#1\ee{\begin{equation}#1\end{equation}}

\begin{document}

\title{Efficient discrimination between real and complex quantum theories}

\author{Josep Batle}
\affiliation{Departament de F\'isica UIB i Institut d'Aplicacions Computacionals de Codi Comunitari (IAC3), Campus UIB, 
E-07122 Palma de Mallorca, Balearic Islands, Spain}
\affiliation{CRISP - Centre de Recerca Independent de sa Pobla, 07420 sa Pobla, Balearic Islands, Spain}
\author{Tomasz Bia{\l}ecki}
\affiliation{Faculty of Physics, University of Warsaw, ul. Pasteura 5, PL02-093 Warsaw, Poland}
\author{Tomasz Rybotycki}
\affiliation{Systems Research Institute, Polish Academy of Sciences, 6 Newelska Street, PL01-447 Warsaw, Poland}
\affiliation{Nicolaus Copernicus Astronomical Center, Polish Academy of Sciences, 18 Bartycka
Street, PL00-716 Warsaw, Poland
}
\affiliation{Center of Excellence in Artificial Intelligence, AGH University,
30 Mickiewicza Lane, PL30-059 Cracow, Poland}
\author{Jakub Tworzyd{\l}o}
\affiliation{Faculty of Physics, University of Warsaw, ul. Pasteura 5, PL02-093 Warsaw, Poland}
\author{Adam Bednorz}
\email{Adam.Bednorz@fuw.edu.pl}
\affiliation{Faculty of Physics, University of Warsaw, ul. Pasteura 5, PL02-093 Warsaw, Poland}

\begin{abstract}

We improve the test to show the impossibility of a quantum theory based on real numbers by a larger ratio of complex-to-real bound on a Bell-type parameter. 
In contrast to previous theoretical and experimental proposals the test requires three settings for the parties $A$ and $C$, but also six settings for the middle party $B$, assuming
separability of the sources. The bound we found for this symmetric configuration imposed on a real theory is $14.69$ while the complex maximum is $18$. This large theoretical difference
enables us to demonstrate the concomitant experimental violation on IBM quantum computer via a designed quantum network, without resorting to error mitigation, obtaining as a result $15.44$ at more than $100$
standard deviations above the found real bound.

\end{abstract}

\maketitle

\section{Introduction}

Quantum mechanics is based on complex numbers from its early days \cite{Heis1,Heis2,Schrod}.
Contrasting real- and complex-based quantum theories may bear little relevance for practical purposes, for it is known in 
several branches of physics that a description based on real numbers alone does not suffice to match experimental results. 
Thus, real and imaginary parts of the wave function are necessary, at least experimentally \cite{realism}. 
It has been pointed out \cite{lit1,lit3,lit4,lit5,lit6,lit7} that one can replace complex with real numbers by doubling the concomitant complex 
$n$-dimensional Hilbert spaces to real-valued ones. 
However, this mathematical equivalence comes at the cost of dealing with extra degeneracy of states, where not all of them are doubled, 
in particular the ground state. This would not entail a problem for local phenomena, but separable states consisting of several parties are doubled in each party. 
Therefore, to reduce the degeneracy one needs extra entanglement in real space, that is, more resources.

Recently, Renou {\it et al} \cite{cvr} developed a test designed whether the states separable in complex space can be replaced by an entangled
state in real space. It is an approach that essentially conjugates several tools borrowed from quantum information theory. 
The real separability imposes additional constraints on correlations, leading to an inequality, with a lower bound for
real states than for complex ones. The test is analogous to the Bell-type tests of nonlocality and local variable models (LVM), involving separated parties. 
In the real-complex test, there are two sources, $P$ and $Q$ and three observers, $A$, $B$, and $C$, where $A$ and $B$ are connected to the source $P$
while $C$ and $B$ to the source $Q$. Then $B$ makes a single measurement with $4$ outcomes, while $A$ and $C$ make dichotomic measurements for 
three and six settings, respectively. The violation of the inequality rules out real separability, which has been verified experimentally, \cite{cvre1,cvre2}. However, the first experiment used photons, which can get lost, so the results were postselected to coincidences. In the second experiment, due to errors, the resulting  correlations have been enhanced by error mitigation \cite{erm,erm2}, multiplying the results by the inverse fidelity matrix. The same applies to a recent IBM Quantum test \cite{cvribm}.

In the same vein as non-locality tests can be challenged by LVM with the introduction of loopholes of different 
nature (locality, detector-efficiency, setting-independence, etc), the experimental tests of real quantum theory are not exempt from them.
The locality loophole has been closed recently \cite{cvre3}, but the efficiency loophole persisted in all previous experiments.
Reduction to four or three settings \cite{cvrab} makes the loophole even harder to close although independence of sources lowers respective real bounds a bit \cite{cvrb}.
In the present contribution, we show that the gap between real and complex theories can be actually widened if one allows six settings for the middle
party $B$ with 4 outcomes. It is allowed, as the parties are expected to be spatially separated. The correlations are generated by permutations of the set $\{1,2,3\}$ corresponding to three settings. We obtained an upper bound (not necessarily tight) analytically and confirmed it by semidefinite programming (SDP). 
On the contrary, the complex quantum bound is 18, obtained by settings of $A$ and $C$ along the main axes of the Bloch sphere, while the $B$ settings are obtained from the maximally entangled state and six rotations of the cube inscribed inside the Bloch sphere. 

As opposed to \cite{cvre1,cvre2}, we considerably change the experimental settings, upon which we certify our theoretical proposal. We also construct new Bell inequalities. 
Furthermore, we perform our experimental validation in a public venue, such as IBM Quantum, making our results highly reproducible without access to complex experimental 
facilities. Notice that some universal gates have to be translated into built-in native gates.

The present contribution is divided as follows. We begin with the description of the setup and the notation. Then we construct the inequality involving six settings of $B$. Finally, we present the demonstration of the 
violation in a quantum network designed on an IBM Quantum device, without error mitigation. Note that the small size of this device makes it impossible to close the locality loophole there.

\section{General setup of the test}

\begin{figure}
\begin{center}
\includegraphics[scale=.45]{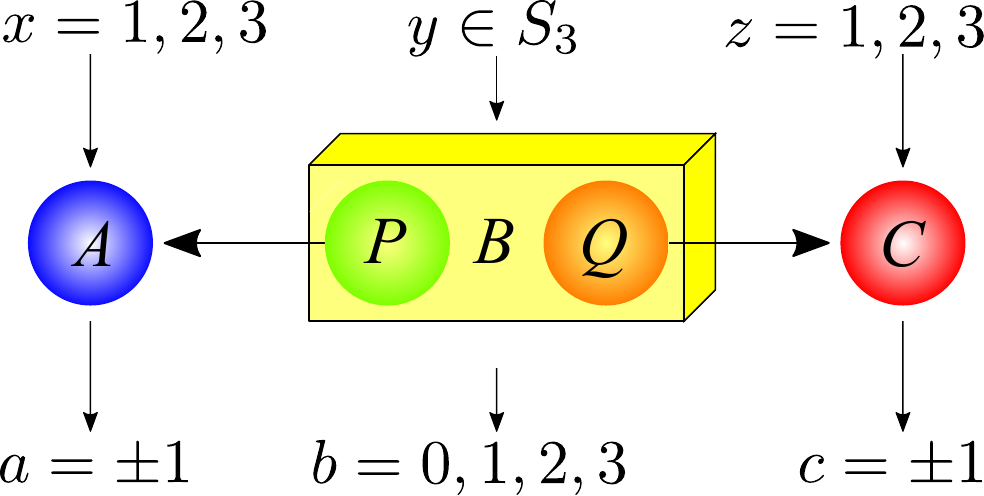}
\end{center}
\caption{ The setup of the test. The separate sources $P$ and $Q$ generate entangled states shared between $AP$ and $QC$ respectively. Central parts $P$, $Q$ of these states are measured jointly by $B$ with four possible outcomes $b$ for settings $\eta$, corresponding to six permutations of the group $S_3$. The left and right parts are measured by observers $A$ and $C$, with dichotomic outcomes $a$ and $c$ for settings $x$ and $z$, respectively.}\label{abc}
\end{figure}

The minimal setting \cite{cvr} has to consist of at least three observers $A$, $B$, and $C$, which we depict in Fig. \ref{abc}.
The sources $P$ and $Q$ are prepared separately from each other, so we assume initially no shared entanglement between $P$ and $Q$. In quantum mechanics based on real numbers, the separability between $P$ and $Q$ leads to tighter correlation bounds
than in the full complex space. We refer
to this property as real separability in the rest of the paper.

The Hilbert space of the model setup can be described as a product of 4 subspaces: $A$, $P$, $Q$, $C$.
The source states $\rho_{AP}$ and $\rho_{QC}$ are entangled in the respective spaces, but they remain separated from each other, i.e.
the initial state is $\rho=\rho_{AP}\otimes\rho_{QC}$. 
In the spirit of making minimum requirements for a practical implementation of a quantum network, we will consider all $A$, $P$, $Q$, $C$ realized by a single qubit each.
Measurement of $B$ in the space of $PQ$ is implemented on two qubits simultaneously, with the outcomes $b=0,1,2,3$ denoting the measurement of $00, 01, 10, 11$ in the computational basis. Measurements $A$ and $C$ are standard measurements of a single qubit.

We assume the condition of separability in a general form
\be
\rho=\sum_\lambda p^\lambda \rho_{AP}^\lambda\otimes\rho_{QC}^\lambda,
\ee
with some unknown probability $p^\lambda\geq 0$, $\sum_\lambda p^\lambda=1$. The states $\rho_{AP}$, $\rho_{QC}$ are expressed in real/complex quantum mechanics when searching for the corresponding bound. The separability is the critical assumption as otherwise an additional entanglement between $P$ and $Q$ would make real and complex descriptions indistinguishable.

We are interested in the correlations of observables measured by $A$, $B$, and $C$,
\be
\langle A_x B_{b\eta} C_z \rangle=\sum_{a,c=\pm 1}ac\;\mathrm{Tr} \left( \rho\,  A_{a|x}\otimes B_{b|\eta}\otimes C_{c|z}  \right) \label{abcpq}.
\ee
We simplify the notation on the left hand side, as the observables $A_x=A_{1|x}-A_{-1|x}$, with Hermitian $A_{a|x}\geq 0$, $\sum_{a}A_{a|x}=1$
($C$ analogously) contribute directly the values $a,c=\pm 1$ into the product, while the observable $B_{b\eta}$ takes values $0,1$ indicated by $b$. The fixed settings for measurements are specified by a tuple $(x,\eta,z)$. The $A/C$ observables have the values $\pm 1$ so $-1\leq A_x,C_z\leq 1$ while $B_{b\eta}\geq 0$ and $\sum_b B_{b\eta}=1$.

The test discriminating real and complex quantum theories will be a linear combination of the above correlations, optimized so that the obtained real bound is much lower than the full complex bound. In previous proposals, the observer $B$ had only one setting and four outcomes \cite{cvr,cvre2,cvre3}.
We conducted an intensive numerical study of a wide range of parameters to see if we could overcome the previous maxima found in \cite{cvr, cvrab}. The corresponding survey involved random uniform exploration of a set of coefficients of the linear functional, while optimizing each quantity with complex or real values. Despite extensive numerical efforts, the difference between the real and complex theory did not improve significantly over the original proposal. With these findings in mind, we turned to the possibility of multiple settings for the party $B$ \cite{cvre1}.

\begin{figure}
\begin{center}
\includegraphics[scale=1]{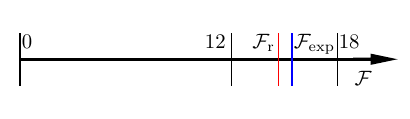}
\end{center}
\caption{The value of the Bell-type parameter $\mathcal F$ for the IBM demonstration, $\mathcal F_{\mathrm{exp}}$ (blue), compared to our real  quantum bound $\mathcal F_{\mathrm{r}}\simeq 14.6969$ (red), classical bound $12$ and full complex quantum bound $18$ (black). The error is below the width of the vertical line.}
\label{crbar}
\end{figure}

\section{Results for six middle party settings}

The observers $A$ and $C$ can choose one of three settings $x=1,2,3$ and $z=1,2,3$, respectively.
We allow the observer $B$ to choose one of six settings enumerated 
by a permutation $\eta\in S_3:\{1,2,3\}\to \{1,2,3\}$, which we denote by a tuple $\eta\equiv \left( \eta(1)\eta(2)\eta(3) \right)$
i.e. $(123)$, $(231)$, $(312)$, $(321)$, $(213)$, $(132)$. 

In Appendix \ref{appa}, we prove that the separability in real space implies that
\be
\mathcal F=\sum_{\eta bz}\mathrm{sgn}\,\eta\,  (-1)^{\delta_{zb}+\delta_{0b}}\langle A_{\eta(z)}B_{b \eta}C_z\rangle\leq \mathcal F_r,
\label{frr}
\ee
where
\be
\mathcal F_{\mathrm r}=6\sqrt{6}\simeq 
14.6969.
\label{frr2}
\ee

This constitutes the main result of our work and is illustrated in Fig. \ref{crbar}.
It is greater than the classical maximum, which is $12$ based on the derivation in the Local Variables Model for $A_x,C_z=\pm 1$ and $B_{b \eta}=0,1$.

The algebraic maximum of $\mathcal F$ is 18 as each permutation and each $z$ can give maximally $1$, since $-1\leq AC\leq 1$ and $0\leq B_{b}\leq 1$ and 
$\sum_b B_{b}=1$.
We will now show that this limit is attainable in a full complex quantum framework. This derivation illustrates the method and gives the notation we use in the Appendix \ref{appa} to determine the real maximum.
By linearity, it suffices to consider pure initial states $\rho_{AP}=|\psi_{AP}\rangle\langle\psi_{AP}|$,
$\rho_{QC}=|\psi_{QC}\rangle\langle\psi_{QC}|$. Our measurement is projective in $PQ$ space, so $B_{b\eta}=|\psi_{b\eta}\rangle\langle\psi_{b\eta}|$.
We assume each party, $A$, $P$, $Q$, $C$ to be a separate two level space with the basis $|0\rangle$, $|1\rangle$.
It is convenient to rewrite (\ref{abcpq}) identifying the general state $|\psi\rangle=\sum_{ij}\Psi_{ij}|ij\rangle$,
with the $2\times 2$ matrix $\Psi$ with entries $\Psi_{ij}$,
in $AP$, $QC$, and $PQ$ spaces
\be
\langle A_x B_{b\eta} C_z\rangle=\mathrm{Tr}\Psi^\dag_{b\eta}\Psi^T_{AP} A_x^T\Psi^\ast_{AP} \Psi_{b\eta} \Psi^\ast_{QC} C_z\Psi^T_{QC}.
\ee
Now let us take $A_j=C_j=\sigma_j$ with Pauli matrices $\sigma_j$ ($j=1,2,3$), and
the initial states represented by $2\times 2$ matrices
$
\Psi_{AP}=\Psi_{QC}=\bar{\Psi}=i\sigma_2/\sqrt{2},
$
corresponding to the singlet Bell state 
\be
|\bar{\psi}\rangle=(|01\rangle-|10\rangle)/\sqrt{2}\label{sbe}.
\ee
The four outcome states are the rotated Bell states $\Psi_{b\eta}=R_\eta\sigma_b\sigma_2/\sqrt{2}$, with the corresponding rotation $R_\eta$ on the Bloch sphere.

The correlation reads
\be
\langle A_x B_{b \eta}C_z\rangle=-\mathrm{Tr}R^\dag_\eta A_xR_\eta\sigma_b C_z\sigma_b/8,
\label{cor}
\ee
since $\sigma_2 A_x^T\sigma_2=-A_x$.
We take $R_\eta$ such that it permutes all $A_x$ observables:
\be
R_\eta(A_1,A_2,A_3)R^\dag_\eta=\mathrm{sgn}\, \eta\, (A_{\eta(1)},A_{\eta(2)},A_{\eta(3)}).
\label{rot}
\ee
Such rotations are, by Pauli matrix algebra,
$
R_{123,231,312}=[(\sigma_0-i\sigma_1-i\sigma_2-i\sigma_3)/2]^{0,1,2}
$, 
i.e. the $2\pi/3$ rotation about the principal diagonal (essentially circulating the directions $123$)
and 
$
R_{132,321,213}=(\sigma_{3,1,2}-\sigma_{2,3,1})/\sqrt{2}
$, 
i.e. $\pi$ rotations about the in-plane diagonals on the Bloch sphere.

The algebra to calculate (\ref{frr}) from (\ref{cor})
uses the identity $\sigma_b\sigma_j\sigma_b=(-1)^{\delta_{jb}+\delta_{0b}+1}\sigma_j$.
All in all, by $\sigma_j^2=1$, each correlation in (\ref{frr}) gives $1/4$ for each $z,\eta,b$ and  $1$ summed over $b$, which is 18 in total.

\section{Implementation on IBM}

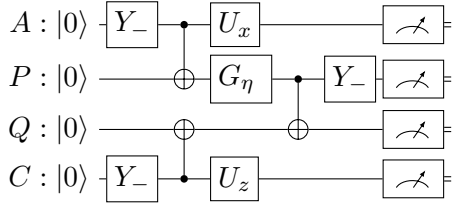
\begin{figure}
\begin{center}
\begin{tikzpicture}[scale=1]
		\begin{yquant*}
			init {$A:\ket 0$} q[0];
			init {$P:\ket 0$} q[1];
			init {$Q:\ket 0$} q[2];
			init {$C:\ket 0$} q[3];
			box {$Y_-$} q[0];
			box {$Y_-$} q[3];
			cnot q[1] | q[0];
			cnot q[2] | q[3];
			box {$U_x$} q[0];
			box {$U_z$} q[3];
			box {$G_\eta$ } q[1];	
			cnot q[2] | q[1];
			box {$Y_-$} q[1];
			
			measure q[0,1,2,3];
		\end{yquant*}

\end{tikzpicture}
\end{center}
\caption{The circuit implementing the complex-real test for the correlation $\langle A_xB_{b\eta }C_z\rangle$. Here $Y_-=Z_-X_+Z_+$ and $U_{1,2,3}=I,\,X_+,\,X_+Z_+$, respectively. The gate $G_\eta$ corresponds to the appropriate permutation, see Fig. \ref{gsig}. The $CNOT_\downarrow|ab\rangle$ gate links the control qubit $a=\bullet$ with  the target qubit $b=\oplus$.}
\label{cvrtest}
\end{figure}

\begin{figure}
\begin{center}
\begin{tikzpicture}[scale=1]
		\begin{yquant*}
			init  {} q[0,1,2,3,4,5];
			box {$X_+$} q[1];
			box {$Z_+$} q[1];
			box {$Z_+$} q[2];
			box {$X_+$} q[2];
			box {$Z$} q[2];
			box {$Z_+$} q[3];
			box {$X_+$} q[4];
			box {$Z_-$} q[5];
			box {$X_+$} q[5];
			box {$Z_+$} q[5];
			output {$123$} q[0];
			output {$231$} q[1];
			output {$312$} q[2];
			output {$132$} q[3];
			output {$213$} q[4];
			output {$321$} q[5];
		\end{yquant*}

\end{tikzpicture}
\end{center}
\caption{The gate  $G_\eta$ for each permutation.}
\label{gsig}
\end{figure}
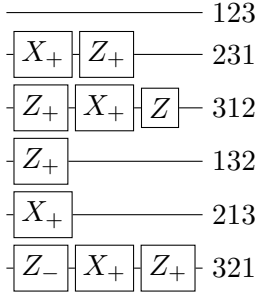

We have demonstrated the above test on IBM Quantum, exceeding our real bound 
as shown in Fig. \ref{crbar}. The implementation uses 4 qubits, corresponding to $APQC$, connected by $CNOT$ gates (or equivalent),
see Fig. \ref{cvrtest}.
The initial state of each of the four is $|0\rangle$. 
The initial state $AP$ and $QC$ is realized by $CNOT$ gates, i.e.
\begin{equation}
CNOT_\downarrow(Y_-I)|00\rangle=(|00\rangle-|11\rangle)/\sqrt{2}=(XI)|\bar{\psi}\rangle,\label{enti}
\end{equation}
for the Bell state (\ref{sbe}).
We use the convention for tensors that $(AB)|ab\rangle$ means $(A|a\rangle)(B|b\rangle)$ and $V_\pm=\exp(\mp i\pi V/4)$ with $V$
being one of Pauli gates $I,X,Y,Z=\sigma_{0,1,2,3}$,
ignoring global phase factors $e^{i\phi}$ (see Appendix \ref{appb} for the explicit notation).
We also denote two-qubits gates by $\downarrow$ and $\uparrow$, which mean the direction of the gate (it is not symmetric), i.e.
$\langle a'b'|G_\uparrow|ab\rangle=\langle b'a'|G_\downarrow|ba\rangle$. 
In particular, in the $|0\rangle$, $|1\rangle$ basis,
\be
Y_-=\frac{1}{\sqrt{2}}
\begin{pmatrix}
1&1\\
-1&1
\end{pmatrix}.
\ee
$CNOT$ toggles the target state $|0\rangle$, $|1\rangle$ on condition that the control state is $|1\rangle$, reading
\be
CNOT_\downarrow=
\begin{pmatrix}
I&0\\
0&X\end{pmatrix}
=\begin{pmatrix}
1&0&0&0\\
0&1&0&0\\
0&0&0&1\\
0&0&1&0\end{pmatrix},
\ee
in the basis $|00\rangle$, $|01\rangle$, $|10\rangle$, $|11\rangle$.
The measurement operators $A_j=U_j^\dag ZU_j=C_j$
with $U_1=I$, $U_2=X_+$, $U_3=X_+Z_+$. Operationally, $Z=|0\rangle\langle 0|-|1\rangle\langle 1|$ is measured as a difference of $|0\rangle$ and $|1\rangle$ occurences.
It gives $A_1=Z$, $A_2=Y$, $A_3=X$. 
The outcome in $PQ$ space is defined
$B_{b\eta}=G^{\prime\dag}_\eta M_{b\eta}G'_\eta$
with
\be
G'_\eta=(Y_-I)CNOT_\downarrow(G_\eta I),
\ee
and $G_{123}=I$, $G_{231}=Z_+X_+$, $G_{312}=ZX_+Z_+$, $G_{132}=Z_+$, $G_{213}=X_+$, $G_{321}=Y_+$, see Fig. \ref{gsig}.
To understand such a choice, note that
\begin{align}
&\langle 00|(Y_-I)CNOT_\downarrow=\nonumber\\
&(\langle 00|+\langle 11|)/\sqrt{2}=\langle\bar{\psi}|(YI),\nonumber\\
&\langle 01|(Y_-I)CNOT_\downarrow=\nonumber\\
&(\langle 01|+\langle 10|)/\sqrt{2}=\langle\bar{\psi}|(ZI),\nonumber\\
&\langle 10|(Y_-I)CNOT_\downarrow=\nonumber\\
&(\langle 11|-\langle 00|)/\sqrt{2}=\langle\bar{\psi}|(XI),\nonumber\\
&\langle 11|(Y_-I)CNOT_\downarrow=\nonumber\\
&(\langle 10|-\langle 01|)/\sqrt{2}=\langle\bar{\psi}|.\label{xyzi}
\end{align}

The explicit correspondence between $G_\eta$ and permutation is
\begin{align}
&G_{231}=(1-iZ-iX-iY)/2,\nonumber\\
&G_{312}=(1+iX+iY+iZ)/2,\nonumber\\
&G_{132}=(1-iZ)/\sqrt{2}=X(X-Y)/\sqrt{2},\nonumber\\
&G_{213}=(1-iX)/\sqrt{2}=Y(Y-Z)/\sqrt{2},\nonumber\\
&G_{321}=(1-iY)/\sqrt{2}=Z(Z-X)/\sqrt{2}.\label{ggg}
\end{align}
The signs in $231$ and $312$ are opposite to the previous section because the orientation $X,Y,Z$ is opposite to $Z,Y,X$.
The projection $M=|m\rangle\langle m|$ given by the $PQ$ two-qubit state is specified in Table \ref{kpq}.
In particular, the identity corresponds to $2,1,3,0$ since the observables $I,Z,Y,X$ appearing in (\ref{xyzi}) correspond to $\sigma_b$ from Sec. 3 with $b=0,1,2,3$.
However, due to the fact that the entangled state (\ref{enti}) is rotated by $X$ with respect to the singlet Bell state,
in practice we deal with $A'_j=XA_jX=C'_j$, which results in $A'_1=-Z$, $A'_2=-Y$, $A'_3=X$.
To take it into account, we can flip the Pauli algebra, using $X'=X$, $Y'=-Y$, $Z'=-Z$, which preserves the commutation relations.
Now the permutations read
\begin{align}
&G_{231}=Z'(1-iZ'-iY'-iX')/2,\nonumber\\
&G_{312}=Y'(1+iX'+iY'+iZ')/2,\nonumber\\
&G_{132}=Y'(Y'-X')/\sqrt{2},\nonumber\\
&G_{213}=Y'(Y'-Z')/\sqrt{2},\nonumber\\
&G_{321}=X'(X'-Z')/\sqrt{2}.\label{gggp}
\end{align}
To map $b$ to the $PQ$ state $m$, we multiply the outcomes by $Z,Y,Y,Y,X$ for respective permutations.
From Pauli algebra $XY=-YZ=iZ$, $YZ=-ZY=iX$, $ZX=-XZ=iY$,
we have  a permutation of outcomes, $0123\to 1032,2301,3210$ for $Z,Y,X$, respectively, resulting in Table \ref{kpq}.

\begin{table}
\begin{center}
\begin{tabular}{c|*{4}{c}}
\toprule
$\eta$&$00$&$01$&$10$&$11$\\
\midrule
$123$&$2$&$1$&$3$&$0$\\
$231$&$3$&$0$&$2$&$1$\\
$312$&$0$&$3$&$1$&$2$\\
$132$&$0$&$3$&$1$&$2$\\
$213$&$0$&$3$&$1$&$2$\\
$321$&$1$&$2$&$0$&$3$\\
\bottomrule
\end{tabular}
\end{center}
\caption{The outcome $b$ depending on the permutations $\eta$ and the states $pq$ of the qubits $P$ and $Q$.}
\label{kpq}
\end{table}

\begin{figure}
\begin{center}
\includegraphics[scale=1.2]{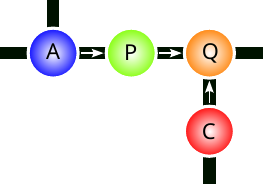}
\end{center}
\caption{The actual topology of qubits used in the test on \emph{ibm\_brisbane}  for the circuit in Fig. \ref{cvrtest}. The black connections indicate 
two-qubit gates linking the test qubits and external ones. The arrows show the
directions of $ECR$ gates between the test qubits (see Appendix \ref{appb}). }
\label{qub}
\end{figure}

\begin{table}
\begin{center}
\begin{tabular}{*{4}{c}}
\toprule
device/qubit & freq. (GHz)&r/a error\\
\midrule
47(A)&4.770&$6.0\cdot 10^{-3}$\\
48(P)&4.844&$1.3\cdot 10^{-2}$\\
49(Q)&4.697&$9.2\cdot 10^{-3}$\\
55(C)&4.837&$11.7\cdot 10^{-3}$\\
\bottomrule
\end{tabular}
\end{center}
\caption{The characteristics of the qubits used in the demonstration,
frequency between 0 and 1 level, readout/assignment error. The duration of the single gate pulse is always 60ns.}
\label{tech}
\end{table}

\begin{figure}
\begin{center}
\includegraphics[scale=.8]{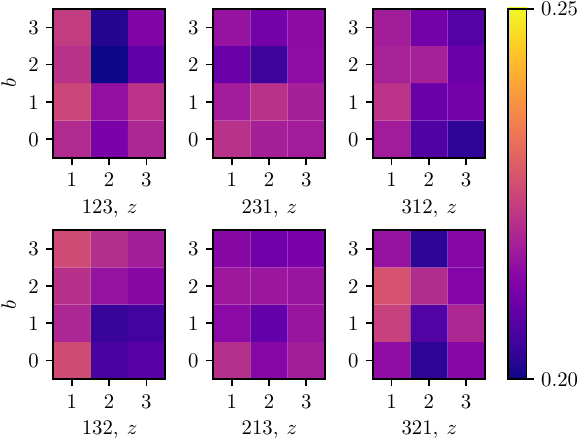}
\end{center}
\caption{The $\mathrm{sgn}\,\eta\, (-1)^{\delta_{jb}+\delta_{0b}}\langle A_{\eta(z)}B_{b\eta}C_z\rangle$ correlations for all values $z$, $b$, $\eta$,
from the IBM Quantum demonstration.
In the ideal case, they are all equal to $0.25$.}
\label{mesh}
\end{figure}
 We performed the test on \emph{ibm\_brisbane}, qubits 47 ($A$),48 ($P$),49 ($Q$),55 ($C$), chosen due to their relatively low errors, with native single qubit gates $X_+$ phase shifts $Z_\pm$ and $CNOT$ gates transpiled by the native $ECR$ gates, see Fig. \ref{qub} and Appendix \ref{appb}.
The errors of $ECR$ gates $48\to 47$, $48\to 49$, $55\to 49$, are $5.2\cdot 10^{-3}$, $1.1\cdot 10^{-2}$, $1.3\cdot 10^{-2}$, respectively.
The single qubit characteristics is specified in Table \ref{tech}.
To determine the total error, we assume independence between experiments (we randomly shuffle them in each job). 
For a given $x$, $\eta$ and $z$ the contribution to the error is
\be
N\langle \Delta \mathcal F^2_{x\eta z}\rangle=1-\mathcal F_{x\eta z}^2,
\ee
where 
\be
\mathcal F_{x\eta z}=\sum_b(-1)^{\delta_{zb}+\delta_{0b}}\langle A_{x}B_{b\eta}C_z\rangle,
\ee
and $x=\eta(z)$,
is the partial witness contribution, omitting global signs. Here $N$ is the total number of trials.
Then
\be
\langle \Delta \mathcal F^2\rangle=\sum_{x\eta z}\langle \Delta \mathcal F_{x\eta z}^2\rangle.
\ee
Using 6 jobs, 20000 shots for all 18 correlations we obtained the value
$\mathcal F =15.4436 \pm 0.0066$ which is above our real threshold by  more than 100 standard deviations.
The scale of the deviation from 18 is roughly consistent with the gate and readout errors.

The individual correlations are presented in Fig. \ref{mesh} while the obtained value related to the other bounds is shown in Fig. \ref{crbar}.
As a sanity test, we checked also no-signaling behavior on the data we collected, i.e. if a party's setting does not affect the results of the other party 
when ignoring its outcomes. We have not found  any significant signaling effects, 
although the times and distances in IBM Quantum devices cannot, in principle, rule it out.
The data and scripts are publicly available \cite{zen}.

\section{Conclusion}

We have shown that the discrimination between real and complex quantum theories can be tested with non-ideal resources using a publicly available quantum computer,
without additional steps such as inverse fidelity matrix.
An open question remains if the test and the real bound we have found are optimal (our computational efforts did not elucidate that question). One can also try to perform the test
at larger distances to close the locality loophole simultaneously. We believe that the answer to the question of whether quantum mechanics needs complex numbers or not is
now undoubtedly answered in the positive.

\section*{Acknowledgments}
We thank M.-O. Renou for inspiring discussions.
 The results have been created using IBM Quantum. The views expressed are those of the authors and do not reflect the official policy or position of IBM or the IBM Quantum team. 
TR gratefully acknowledges the funding support by the
program ,,Excellence initiative research university'' for the AGH University in
Krakow as well as the ARTIQ project: UMO-2021/01/2/ST6/00004 and
ARTIQ/0004/2021.
\appendix
\section{Derivation of an upper bound for real separable states}
\label{appa}

We shall derive the upper bound given by Eqs. (\ref{frr}) and (\ref{frr2}). 
The permutation sign reads explicitly
\be
\mathrm{sgn}\;\eta=\left\{\begin{array}{ll}
+1&\mbox{ for }\eta=123,231,312,\\
-1&\mbox{ for }\eta=321,132,213.\end{array}\right.
\ee
By linearity, we can consider only operators satisfying $A^2=C^2=1$. Let us introduce the following abbreviation
\be
i\equiv (-1)^{\delta_{ib}+\delta_{0b}}\mathrm{sgn}\;\eta\: A_{\eta(i)}C_{i},
\ee
for $i=1,2,3$.
To avoid ambiguity, we shall write products of the above operator factors in square brackets $[\cdots ]$ except empty bracket $[]\equiv 1$ (identity).
Note that $[\ast ii\star]=[\ast\star]$.
Now, we construct the nonnegative expression
\be
\sum_{b\eta}\langle G_{b\eta}  B_{b\eta}\rangle\geq 0,
\ee
with  $G_{b\eta}$ equal to a sum of squares. To find the best sum of squares, we have considered a generic multivariate family of squares, and optimized it, supported by Mathematica and Matlab optimization, and guessing analytic reductions by examining the results. We have found the sum of the following squares
\be
(x-[1]-[2]-[3])^2+\sum_{ijk\in Z_3}(t+t[j]-([i]+[jk])/2t)^2,
\ee
with free real parameters $t$ and $x$
for cyclic shifts $Z_3=123,231,321$.
Here $0\leq B_{b\eta}\leq 1$, for real symmetric matrices $B_{b\eta}$, but $\sum_b B_{b\eta}=1$, while the square for nonsymmetric matrices should be operationally understood as
$O^2=O^TO$, with $O^T$ meaning the transpose (reversed order in products) of $O$, since the positivity applies to arbitrary vectors $|v\rangle$,  $\langle Ov|Ov\rangle=\langle v|O^TO|v\rangle \geq 0$. 

Then
\begin{align}
&G_{b\eta}=
3+x^2+6t^2+3/2t^2\nonumber\\
&-2(x+1-t^2)([1]+[2]+[3])+\sum_{ijk\in S_3}[ijk]/4t^2,
\end{align}
for permutations $S_3=123$,$231$,$312$,$321$,$132$,$213$,
as $[ij]$ terms cancel.
The Bell type parameter reads 
\be
\mathcal F=\sum_{b\eta} \langle ([1]+[2]+[3])B_{b\eta}\rangle.\label{fbb}
\ee
The term $[ijk]$ is independent of $b$ as the sign exponent $\delta_{ib}+\delta_{jb}+\delta_{kb}+3\delta_{0b}=1+2\delta_{0b}$, and the sign of $\eta$
appears three times. The real separability implies equality with partial transpose, i.e. the $AC$ density matrix satisfies $\rho_{aa'cc'}=\rho_{a'acc'}$,
and so the term $[ijk]$ cancels with its transpose due to the opposite sign of the permutation. For instance
\ba
&-\sum_{b}\langle [123]B_{b,123}\rangle=\langle A_{1}C_1A_2C_2A_3C_3\rangle=\nonumber\\
&=\langle A_1A_2A_3C_1C_2C_3\rangle=\langle A_3A_2A_1C_1C_2C_3\rangle=\nonumber\\
&= \langle A_3C_1A_2C_2A_1C_3\rangle=\sum_{b}\langle [123]B_{b,321}\rangle,
\ea
where the middle equality follows from the real partial transpose.
To get the best constraint we have to minimize the bound on (\ref{fbb}), i.e.
\be
\mathcal F_{\mathrm r}=3\frac{3+x^2+6t^2+3/2t^2}{x+1-t^2},
\ee
over  $x\geq t^2-1$.
The analysis of the extrema gives the minimum for
\be
t^2=\frac{15+9x}{2x^3+10x^2+6x-6},
\ee
with $x$ being the real root of
\be
x^3+x^2-5x-9=0,
\ee
and
\ba
&\mathcal F^{(2)}_{\mathrm r}=6\frac{2x^3+x^2+9}{(x-1)(3x+5)}=6\frac{2x^3+x^2+9+(x^3+x^2-5x-9)}{(x-1)(3x+5)}\nonumber\\
&=6x=2\left(\sqrt[3]{98+18\sqrt{17}}+\sqrt[3]{98-18\sqrt{17}}-1\right)\nonumber\\
&\simeq 14.87889449253087
\ea
by Cardano formula, see also the derivation by Mathematica \cite{zen}. We additionally confirmed the bound by SDP code (see \cite{zen,sdp}), by examining formal sums of squares
of expressions containing products up to 2 of each of observables $A$ and $C$, i.e. $1$, $A_x$, $C_z$, $A_xA_{x'}$, $C_zC_{z'}$, $A_xC_z$,
$A_xA_{x'}C_z$, $A_xC_zC_{z'}$, and $A_xA_{x'}C_zC_{z'}$. 
The notation $\mathcal F_r^{(k)}$ corresponds to considering product of maximally $k$ operators $A$ and $C$ inside the quadratic forms.

The bound can be slightly lowered by taking a more complicated expression, i.e. $G_{b\eta}$ as the sum of the following squares, found by numerical and analytical search,
\begin{align}
&(\sqrt{6}-[1]-[2]-[3])^2/11+\sum_{ijk\in Z_3}\nonumber\\
&
\{(36-14\sqrt{6}[i]+3([ij]+[ik])\nonumber\\
&-11([kj]+[jk]))^2/11\nonumber\\
&
+3([ij]-[ik]+[kj]-[jk])^2\}/2^5 3^3\nonumber\\
&+\sum_{ijk\in S_3} (\sqrt{6}[ki]-[iji]-3[k]+2[j])^2/2^4 3^3.
\end{align}
Expanding each of the above squares, we get
\begin{align}
&(\sqrt{6}-[1]-[2]-[3])^2=\nonumber\\
&9-2\sqrt{6}([1]+[2]+[3])+\sum_{i\neq j}[ij],
\end{align}
\begin{align}
&(36-14\sqrt{6}[i]+3([ij]+[ik])-11([kj]+[jk]))^2\nonumber\\
&=2732-2^2 21\sqrt{6}(12[i]+[j]+[k])+\nonumber\\
&2^2 3^3([ij]+[ik]+[ji]+[ki])\nonumber\\
&-2^3 3^2 11([kj]+[jk])\nonumber\\
&+11\cdot 14\sqrt{6}([ikj]+[ijk]+[kji]+[jki])\nonumber\\
&
-33([jikj]+[jijk]+[kikj]+[kijk]\nonumber\\
&+[jkij]+[kjij]+[jkik]+[kjik])\nonumber\\
&
+9([jk]+[kj])+11^2([kjkj]+[jkjk]),
\end{align}
\begin{align}
&([ij]-[ik]+[kj]-[jk])^2=\nonumber\\
&4-[jk]-[kj]-[jkjk]-[kjkj]+\nonumber\\
&[jikj]-[kikj]-[jijk]+[kijk]\nonumber\\
& +[jkij]-[jkik]-[kjij]+[kjik],
\end{align}
\begin{align}
&(\sqrt{6}[ki]-[iji]-3[k]+2[j])^2=\nonumber\\
&20-\sqrt{6}([ijiki]+[ikiji])-6\sqrt{6}[i]\nonumber\\
&
+2\sqrt{6}([ikj]+[jki])+3([ijik]+[kiji])\nonumber\\
&-2([ijij]+[jiji])-6([kj]+[jk]).
\end{align}
Summing up and taking into account permutation symmetry, we 
can identify the terms up to permutations, $1$, $[i]$, $[ij]$, $[iji]$, $[ijk]$, $[ijij]$, $[ijki]$, $[ijik]$, $[kiji]$, $[ijiki]$,
getting
\ba
&2-([1]+[2]+[3])\sqrt{2/3}+\nonumber\\
&\sum_{ijk\in S_3}(9[ijk]-[ijiki])/6^{5/2}.
\ea
The term $[ijiki]$ is also independent of $b$ as the sign depends on
\be
3\delta_{ib}+\delta_{jb}+\delta_{kb}+5\delta_{0b}=1+2\delta_{ib}+4\delta_{0b},
\ee
as an odd product, and cancels with its transpose due to the opposite permutation sign.
Finally, we get
\be
\mathcal F^{(3)}_{\mathrm r}=6\cdot 2\sqrt{3/2}=6\sqrt{6}\simeq 
14.696938456699067.
\ee
We have also confirmed this result by Mathematica and SDP with Matlab, taking into account the cancellation of $[ijk]$, $[ijiki]$ terms \cite{zen}.
The numerical approach showed that more analogous terms, e.g. $[kjkijkj]$
in $\mathcal F_r^{(4)}$, do not improve the bound.

\section{Gates  and transpiling at IBM Quantum}
\label{appb}

We shall use Pauli matrices in the basis $|0\rangle$, $|1\rangle$,
\begin{align}
&X=\begin{pmatrix}
0&1\\
1&0\end{pmatrix},\:Y=\begin{pmatrix}
0&-i\\
i&0\end{pmatrix},\nonumber\\
&\:Z=\begin{pmatrix}
1&0\\
0&-1\end{pmatrix},\:
I=\begin{pmatrix}
1&0\\
0&1\end{pmatrix}.\label{pauli}
\end{align}
The IBM Quantum devices (\emph{ibm\_brisbane}) use transmon qubits \cite{transmon} with the native single qubit gates $X$ and 
\be
X_+=X_{\pi/2}=(I-iX)/\sqrt{2}=\begin{pmatrix}
1&-i\\
-i&1\end{pmatrix}/\sqrt{2},
\ee
denoting $V_\theta=\exp(-i\theta V/2)=\cos(\theta/2)-iV\sin(\theta/2)$ and $V_\pm= V_{\pm \pi/2}$, whenever $V^2$ is identity $I$ or $II$.
Note that $Z_\theta=\exp(-i\theta Z/2)=\mathrm{diag}(e^{-i\theta/2},e^{i\theta/2})$ is a virtual gate adding essentially the phase  shift to next gates.
\cite{zgates}

\begin{figure}[t]
\begin{center}
\begin{tikzpicture}[scale=1]
		\begin{yquantgroup}
			\registers{
			qubit {} q[2];
			}
			\circuit{
			box {$X$}  q[0];
			box {$X_+$}   q[1];
			box {$\downarrow$}  (q[0,1]);
			box {$Z_+$}  q[0];
			}
			\equals
			\circuit{
			cnot q[1] | q[0];
			}
		\end{yquantgroup}
\end{tikzpicture}
\end{center}

\caption{The $CNOT_\downarrow$ gate expressed by $ECR_\downarrow$. }
\label{cnot}
\end{figure}
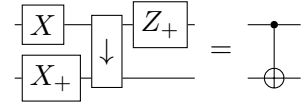

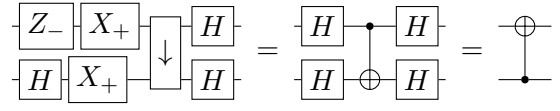
\begin{figure}[t]
 \begin{center}
\begin{tikzpicture}[scale=1]
		\begin{yquantgroup}
			\registers{
			qubit {} q[2];
			}
			\circuit{
			box {$Z_-$}  q[0];
			box {$H$}   q[1];
			box {$X_+$}  q[0];
			box {$X_+$}   q[1];
			box {$\downarrow$}  (q[0,1]);
			box {$H$}  q[0];
			box {$H$}  q[1];
			}
			\equals
			\circuit{
			box {$H$}  q[0];
			box {$H$}   q[1];
			cnot q[1] | q[0];
			box {$H$}  q[0];
			box {$H$}   q[1];
			}
			\equals
			\circuit{
			cnot q[0] | q[1];
			}
		\end{yquantgroup}
\end{tikzpicture}
\end{center}

\caption{The $CNOT_\uparrow$ gate expressed by $ECR_\downarrow$. }
\label{cnotr}
\end{figure}

A native two-qubit gate in  current IBM Quantum  devices is Echoed Cross Resonance ($ECR$)
\cite{ecr} instead of $CNOT$. One can transpile the latter by the former, by adding single qubits gates.
The ECR gate acts on the states $|ab\rangle$ as (Fig. \ref{ecr})
\ba
&ECR_\downarrow=\nonumber\\
&((XI)-(YX))/\sqrt{2}=CR^- (XI) CR^+=\nonumber\\
&
\begin{pmatrix}
0&X_-\\
X_+&0\end{pmatrix}
=\begin{pmatrix}
0&0&1&i\\
0&0&i&1\\
1&-i&0&0\\
-i&1&0&0\end{pmatrix}/\sqrt{2},
\ea
in the basis $|00\rangle$, $|01\rangle$, $|10\rangle$, $|11\rangle$,
with Crossed Resonance gates
\be
CR^\pm=(ZX)_{\pm \pi/4}.
\ee
The gate is its inverse, i.e. $ECR_\downarrow ECR_\downarrow=(II)$.
It can be reversed, i.e., for $a\leftrightarrow b$, we have (Fig. \ref{ecrr})
\ba
&
ECR_\uparrow=((IX)-(XY))/\sqrt{2}=\nonumber\\
&(HH) ECR_\downarrow(Y_+Y_-),
\ea
with Hadamard gate 
\be
H=(Z+X)/\sqrt{2}
=Z_+X_+Z_+=\begin{pmatrix}
1&1\\
1&-1\end{pmatrix}/\sqrt{2},
\ee
$Z_\pm X_+Z_\mp=Y_\pm$, $Y_+=HZ$, and $Y_-=ZH$.
The CNOT gate can be expressed by $ECR$ (Fig. \ref{cnot}),
\begin{align}
&CNOT_\downarrow=((II)+(ZI)+(IX)-(ZX))/2=\nonumber\\
&
\begin{pmatrix}
I&0\\
0&X\end{pmatrix}
=\begin{pmatrix}
1&0&0&0\\
0&1&0&0\\
0&0&0&1\\
0&0&1&0\end{pmatrix}\nonumber\\
&=
(Z_+ I)ECR_\downarrow (XX_+),
\end{align}
while its reverse reads (Fig. \ref{cnotr})
\ba
&CNOT_\uparrow=((II)+(IZ)+(XI)-(XZ))/2
=\nonumber\\
&\begin{pmatrix}
1&0&0&0\\
0&0&0&1\\
0&0&1&0\\
0&1&0&0\end{pmatrix}=\nonumber\\
&
(HH)CNOT_\downarrow(HH)=\nonumber\\
&
(HH)ECR_\downarrow (X_+X_+)(Z_-H).
\ea

\begin{figure}
\begin{center}
\begin{tikzpicture}[scale=1]
		\begin{yquantgroup}
			\registers{
			qubit {} q[2];
			}
			\circuit{
			init {$a$} q[0];
			init {$b$} q[1];
			box {$\downarrow$}  (q[0,1]);}
			\equals
			\circuit{
			box {\rotatebox{90}{$CR^+$}}  (q[0,1]);
			box {$X$} q[0];
			box {\rotatebox{90}{$CR^-$}}   (q[0,1]);}
		\end{yquantgroup}
\end{tikzpicture}
\end{center}

\caption{The notation of the $ECR$ gate in the convention $ECR_\downarrow|ab\rangle$.}
\label{ecr}
\end{figure}
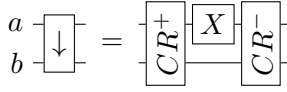
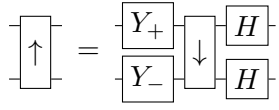
\begin{figure}
\begin{center}
\begin{tikzpicture}[scale=1]
		\begin{yquantgroup}
			\registers{
			qubit {} q[2];
			}
			\circuit{
			box {$\uparrow$} (q[0,1]);
			}
			\equals
			\circuit{
			box {$Y_+$}  q[0];
			box {$Y_-$}   q[1];
			box {$\downarrow$}  (q[0,1]);
			box {$H$}  q[0];
			box {$H$}   q[1];
			}
		\end{yquantgroup}
\end{tikzpicture}
\end{center}
\caption{The $ECR_\uparrow$ gate expressed by $ECR_\downarrow$. }
\label{ecrr}
\end{figure}

\bibliographystyle{quantum}

\end{document}